\documentclass[nofootinbib,floatfix,prd,preprintnumbers,superscriptaddress,twocolumn]{revtex4}
\usepackage{graphicx}
\usepackage{amsfonts}
\usepackage{amssymb}
\usepackage{amsbsy}
\usepackage{amsmath}
\usepackage{latexsym}
\usepackage{bm}
\usepackage{hyperref}

\def\be{\begin{equation}}
\def\ee{\end{equation}}
\def\bea{\begin{eqnarray}}
\def\eea{\end{eqnarray}}

\begin{document}

\title{Self-Accelerating Universe in Galileon Cosmology}

\author{Fabio P Silva}
\affiliation{Institute of Cosmology \& Gravitation,
University of Portsmouth,
Portsmouth~PO1~3FX, United Kingdom\\}

\author{Kazuya Koyama}
\affiliation{Institute of Cosmology \& Gravitation,
University of Portsmouth,
Portsmouth~PO1~3FX, United Kingdom\\}
\date{\today}

\begin{abstract}
We present a cosmological model with a solution that self-accelerates at late-times without signs of ghost instabilities on small scales. The model is a natural extension of the Brans-Dicke (BD) theory including a non-linear derivative
interaction, which appears in a theory with the Galilean shift symmetry. The existence of the self-accelerating universe requires a negative BD parameter but, thanks to the non-linear term, small fluctuations around the solution are stable on small scales. General relativity is recovered at early times and on small scales by this non-linear  interaction via the Vainshtein mechanism. At late time, gravity is strongly modified and the background cosmology shows a phantom-like behaviour and the growth rate of structure formation is enhanced. Thus this model leaves distinct signatures in cosmological observations and it can be distinguished from standard $\Lambda$CDM cosmology.
\end{abstract}

\maketitle

{\bf I. Introduction}

	Observational evidence indicates that the universe is undergoing a period of accelerated expansion \cite{Riess:1998cb,Perlmutter:1998np,Tegmark:2006az,Komatsu:2008hk}. This late-time acceleration is one of the biggest open issues in cosmology today. This would mean that 70\% of the energy content of the universe is dominated by unknown dark energy. The best candidate is vacuum energy, but it is more than 50 orders
of magnitude larger than the required value of the cosmological constant to explain the current acceleration. Alternative explanations are therefore actively being pursued. \\
\indent Modified Gravity is an alternative that interprets this acceleration as a sign that our knowledge of how gravity works at large distances is incomplete and tries to explain the acceleration by a modification of General
Relativity (GR). One of the most studied attempts, the Dvali-Gabadadze-Porrati (DGP) braneworld model \cite{Dvali:2000rv,Dvali:2000hr}, has a self-accelerating solution, i.e. a solution in which cosmic acceleration arises even when no matter is present in our universe \cite{Deffayet:2000uy}. Unfortunately this solution was found to be plagued by ghost instabilities that cast doubt on the validity of the model
(see \cite{Koyama:2007za} for a review and references therein). Other similar alternatives were found (see e.g. \cite{Charmousis:2007ji,deRham:2006pe}) but it seems incredibly difficult to separate the presence of the ghost from the acceleration \cite{Koyama:2009cm}.\\
\indent Recently, an infrared modification of gravity was proposed, which is a generalization of the 4D effective theory in the DGP model \cite{Nicolis:2008in} (see also \cite{Babichev:2009ee}). A novel feature of this effective theory is that the theory is invariant under the Galilean shift symmetry, the constant shift of the gradient of the scalar field   $\partial_{\mu} \phi \to \partial_{\mu} \phi + c_{\mu}$, which keeps the equation of motion at second order \cite{Lovelock}. Then they found self-accelerating de Sitter solutions with no ghost-like instabilities. However, their analysis is valid only for weak gravity in flat space-time and this result might change when the model is covariantized \cite{Deffayet:2009wt,Chow:2009fm,Deffayet:2009mn}. In fact it was shown that the Galilean symmetry cannot be preserved once the theory is covariantized but it is still possible to keep its desired properties. For example, the equation of motion for the scalar field can remain of second order, which is essential because the higher derivative theories contain extra degrees of freedom that are usually related to instabilities. \\
\indent
In this letter, we demonstrate that it is possible to construct a covariant model that keeps desired properties of the Galileon model, i.e. the existence of self-accelerating universe with no ghost-like instabilities on small scales.
We consider the following action (see also \cite{Chow:2009fm})
\be
	\int d^4x\sqrt{-g} \left[ \phi R - \frac{\omega}{\phi}(D\phi)^2 + c \phi + f(\phi) \square\phi (D\phi)^2  + {\cal L}_m \right],
\ee
where $(D\phi)^2 = D^\alpha \phi D_\alpha \phi$ and ${\cal L}_m$ is the matter lagrangian. The cubic interaction is the unique form of interactions at cubic order that keeps the field equation for $\phi$ of second-order \cite{Nicolis:2008in}. The Einstein and field equations are given by
\begin{widetext}
\begin{align}
&G_{\mu \nu} = \frac{1}{2} c g_{\mu \nu} + \frac{1}{\phi} \left( D_\mu D_\nu \phi - g_{\mu \nu} \square\phi \right) + \frac{\omega}{\phi^2} \left( D_\mu \phi D_\nu \phi - \frac{1}{2} g_{\mu \nu} (D\phi)^2  \right) \nonumber\\
\label{eq:ein}\
&\quad\quad\quad - \frac{1}{\phi} \left( \frac{1}{2} g_{\mu \nu} D_\alpha\left[f(\phi)(D\phi)^2 \right] D^\alpha \phi - D_\mu \left[f(\phi) (D\phi)^2 \right] D_\nu \phi + f(\phi) (D_\mu \phi) (D_\nu \phi) \square \phi \right)
+\frac{T_{\mu \nu}}{\phi}, \\
&\frac{3+2\omega}{\phi} \square \phi - c + 2 f(\phi)\Big[(D_\mu D_\nu \phi)(D^\mu D^\nu \phi) - (\square \phi)^2 + R_{\mu \nu} D^\mu \phi D^\nu \phi \Big]\nonumber\\
 \label{eq:eom}
&\quad\quad\quad+ \frac{1}{\phi} \Big[ D_\mu \left[ f(\phi) (D\phi)^2\right] D^\mu \phi + f(\phi) \square \phi (D\phi)^2\Big] + f''(\phi) (D\phi)^4 + 4 f'(\phi)D_\mu \phi D_\nu \phi D^\mu D^\nu \phi =\frac{T^{\mu}_{\mu}}{\phi},
\end{align}
\end{widetext}
where $T_{\mu \nu}$ is the energy-momentum tensor for matter.
\\
 
{\bf II. Background Cosmology}

We found that the simplest choice of $f$ to obtain the self-accelerated universe is given by $f(\phi) = 1/M^2\phi^2$, where $M$ is the parameter of the model. We can also redefine $c$ as $c = -2\Lambda$ so that
it acts like a cosmological constant. For Friedmann-Robertson-Walker spacetime, the Einstein and field equations, (\ref{eq:ein}) and (\ref{eq:eom}) give
\begin{widetext}
\begin{align}
	\label{eq:fried}
	3H^2 &= -3HP + \frac{1}{M^2}\Big[ 3HP^3 + P^4 \Big] + \frac{\omega}{2}P^2 + \Lambda +\frac{\rho}{\phi}, \\
	\label{eq:einB}
	2 \dot H + 3H^2& = - 2HP -\dot P -P^2 + \frac{1}{M^2}P^2\dot P - \frac{\omega}{2}P^2 + \Lambda - \frac{p}{\phi}, \\
\left( 1+\frac{2\omega}{3}\right)\left( \dot P + P^2 + 3 H P \right) &- \frac{2}{3M^2}\Big[ -6HP\left( \dot P + P^2 \right) -9H^2P^2 - 3\dot H P^2 - \frac{5}{2}\left(\dot P+P^2\right)P^2 + \frac{3}{2}HP^3 + 2P^4 \Big] \nonumber\\
	\label{eq:eomB}
	&= \frac{2}{3}\Lambda + \frac{\rho-3p}{3\phi},
\end{align}
\end{widetext}
where $H$ is the Hubble parameter, $\rho$ is the energy density, $p$ is the pressure and $P \equiv \dot{\phi}/\phi $.
\\
\noindent
{\it Self-Accelerating solutions}: A self-accelerating solution can be found by putting $\Lambda=\rho=p=0$ and searching for a solution with $\dot H= \dot P=0$. From Eq.~(\ref{eq:einB}) one can easily show that this solution must satisfy
\be
	\label{eq:aleph}
	\aleph \equiv \frac{P}{H} = \left( -1 \pm \sqrt{-\frac{3\omega}{2}-2} \right) \frac{1}{1+\frac{\omega}{2}}.
\ee

For this solution to exist, the BD parameter must satisfy $\omega <-4/3$. The Friedmann and field equations now yield
\be
	\label{eq:sa}
	H^2 = M^2 \frac{3\left( 1+ \aleph - \omega \aleph^2/6 \right)}{\aleph^3\left( 3+\aleph \right)}.
\ee

\begin{figure}[h]
	\centerline{
		\includegraphics[width=0.5\textwidth]{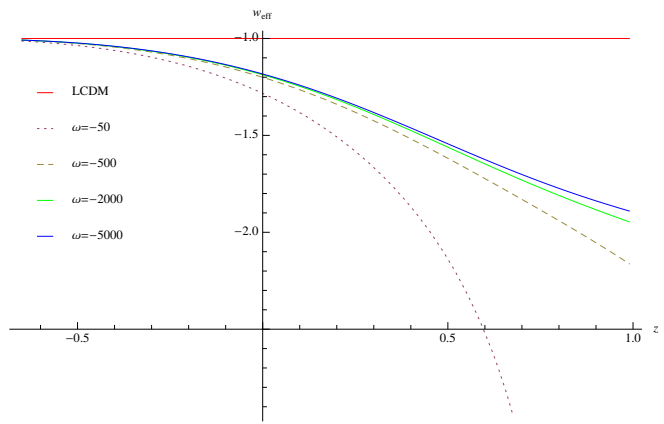}
	}
	\caption{The effective equation of state $w_{\rm eff}$ as a function of redshift for various values of the Brans-Dicke parameter.}
	\label{fig:wDE}
\end{figure}

\begin{figure}[h]
	\centerline{
		\includegraphics[width=0.5\textwidth]{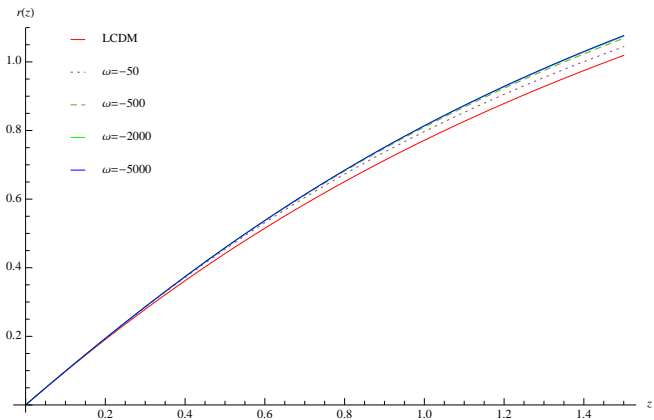}
	}
	\caption{The comoving distance $r(z)$ as a function of redshift for different values of $\omega$.}
	\label{fig:dl}
\end{figure}

This is a self-accelerating de Sitter solution in which the linear terms are equated by the non-linear terms coming from the cubic interaction in Eq.~(\ref{eq:eomB}). Given that $H^2$ has to be positive, we have to pick the negative sign branch in Eq.~(\ref{eq:aleph}). In order to describe the acceleration today, $M$ should be fine-tuned $M \sim H_0$ where $H_0$ is the present-day Hubble parameter.\\
\noindent
{\it Early-time solutions}:
To recover GR at early times $P$ should be smaller than $H$. In fact, GR is recovered if $P=0$. This is
realized at high energies when the non-linear terms in the field equation dominate over the linear term. In this case, from Eqs.~(\ref{eq:fried}-\ref{eq:eomB}), we get $P = M/\sqrt{3} \sim H_0 \ll H$ and we indeed recover the GR solution. This is a cosmological version of the Vainshtein mechanism to screen the scalar field \cite{Chow:2009fm}.\\
\noindent
{\it Numerical solutions}:
We can solve Eqs.~(\ref{eq:einB}-\ref{eq:eomB}) numerically with appropriate initial conditions. Throughout the paper, a flat cosmology is assumed but the inclusion of spatial curvature is straightforward.
We set initial conditions at matter domination era and found the appropriate set of initial conditions in order to get acceleration at present time (defined by $\Omega_m(t_0) = \rho(t_0)/3H(t_0)^2 \phi(t_0)= 0.3$ as an example). Note that Newton's constant is given by $8\pi G = 1/\phi(t_i)$ in this model where $t_i$ is the initial time in the matter dominated era. We obtained cosmological solutions that mimic $\Lambda$CDM, i.e. the scalar field becomes important at late-times and the universe enters a period of accelerated expansion, approaching the self-accelerating solution. We also observed that whenever $P(t_i)$ is appropriately small, $P(t_i) \ll H(t_i)$ the result is insensitive to the initial value of $P(t_i)$. In order to have $\Omega_m(t_0)=0.3$ today, the parameter $M$ must be fine tuned as $M =0.15 H_0$ for $\omega=-50$ and $M=0.024 H_0$ for $\omega=-500$.
The contributions of $P$ in Eq.~(\ref{eq:fried}) can be combined into an effective dark energy, such that the Friedmann equation takes the form $H^2= (8 \pi G) \left( \rho + \rho_{\rm eff} \right)/3$. This allows us to
plot the effective equation of state, $w_{\rm eff}$, of this effective dark energy for different values of the BD parameter as a function of redshift (Fig.~\ref{fig:wDE}).
We also computed the comoving distance $r(z)=\int^z dz'/H(z')$, which is plotted in Fig.~\ref{fig:dl}.
Due the phantom-like behaviour $w_{\rm eff} <-1$, the distance is larger than in $\Lambda$CDM.
\\

{\bf III. Perturbations}\\
\noindent
{\it Stability}:
We now study the stability of the model against small perturbations $\phi \rightarrow \phi(1 + \varphi)$. To do this we expand the field equation (\ref{eq:eomB}) at linear order. We consider perturbations on small scales where we can neglect the expansion of the universe and the effect of metric perturbations. The evolution equation for $\varphi$ becomes
\be
	\label{eq:field}
	d_t(t) \ddot \varphi + d_x(t) \nabla^2 \varphi = 0,
\ee
where
\bea
	\label{eq:time}
	d_t(t) &=& 3+2\omega + \frac{1}{M^2} \Big[ 12H\frac{\dot \phi}{\phi} + 2 \frac{\dot \phi^2}{\phi^2}  +3 \frac{\dot \phi^4}{M^2 \phi^4} \Big], \\
	\label{eq:spatial}
	d_x(t) &=& 3+2\omega + \frac{1}{M^2} \Big[ 4\frac{\ddot \phi}{\phi} + 8H\frac{\dot \phi}{\phi} - 2 \frac{\dot \phi^2}{\phi^2} - \frac{\dot \phi^4}{M^2 \phi^4} \Big]. \nonumber \\
	&&
\eea

\begin{figure}[h]
	\centerline{
		\includegraphics[width=0.5\textwidth]{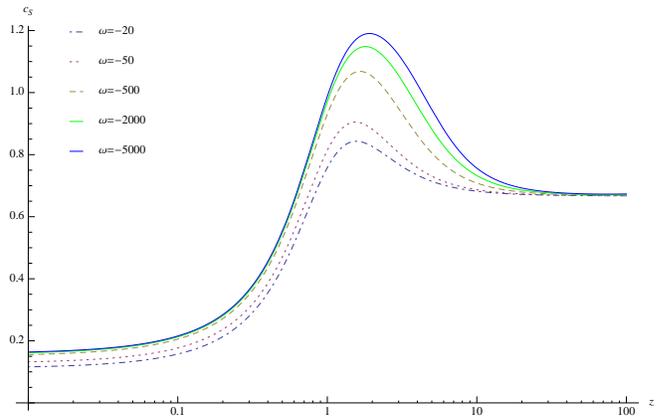}
	}
	\caption{Plot of $c_s$ as a function of redshift for various values of the Brans-Dicke parameter. As can be seen during the transition to the accelerating phase the perturbations go superluminal for $\omega \lesssim -190$.}
	\label{fig:cs}
\end{figure}

In the usual BD theory, if the BD parameter is smaller than $-3/2$, then $d_t$ is negative and $\varphi$ becomes a ghost. In our model, the non-linear interaction term changes the sign of $d_t$. We found that as long as $P$ is positive, $d_t$ and $d_x$ are positive at all times. This can be realized if $\omega <-2$ (see Eq.~(\ref{eq:aleph})). One can also calculate the sound speed of these perturbations, $c_s^2 = d_x/d_t$. At early times, the terms proportional to $H$ dominate and thus $c_s^2 = 2/3$, which is subluminal. The behaviour of $c_s(t)$ for the background solutions discussed above can be found in Fig.~\ref{fig:cs}. The fluctuations can be superluminal for large $|\omega|$ at the transition from the matter dominated era to the accelerating phase and this would impose an upper bound for $|\omega|$.

\noindent
{\it Growth Factor}:
The evolution equation for the cold dark matter over-density $\delta$ is given by:
\be
	\label{eq:growth}
	\ddot \delta + 2 H \dot \delta = \frac{\nabla^2}{a^2} \Psi,
\ee	
where the perturbed line element is given by $ds^2=-(1+2 \Psi)dt^2+a(t)^2 (1+2 \Phi) \delta_{ij}
dx^i dx^j$. On small scales, the Einstein equations (\ref{eq:ein}) give
\bea
	\label{eq:spat}
	\Phi + \Psi &=& -\varphi, \\
	\label{eq:Poisson}
	\frac{\nabla^2}{a^2} \Phi &=& -\frac{1}{2 \phi}\rho \delta  - \frac{1}{2}\left( 1-\frac{P^2}{M^2} \right) \frac{\nabla^2}{a^2} \varphi.
\eea

These can be combined with the field equation (from (\ref{eq:field}))
\be
	d_x(t) \frac{\nabla^2}{a^2} \varphi = - \frac{1}{\phi}\left( 1 + \frac{P^2}{M^2} \right) \rho \delta ,
\ee
where we use the quasi-static approximations and neglected the time derivative term. Then the equation for
$\delta$ is obtained as
\bea
	\label{eq:evo}
	\ddot \delta + 2 H \dot \delta &=& 4 \pi G_{\rm eff} \rho \delta , \\
4 \pi G_{\rm eff} &=&  \left[1 + \left(1 + \frac{P^2}{M^2} \right)^2\frac{1}{d_x} \right] \frac{1}{2\phi}.
\eea

\begin{figure}[h]
	\centerline{
		\includegraphics[width=0.5\textwidth]{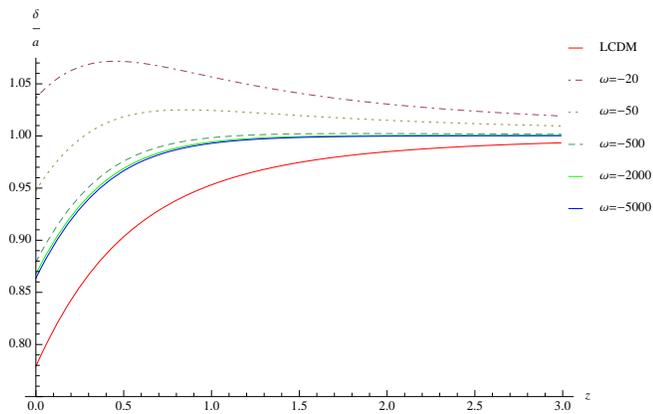}
	}
	\caption{The growth rate $\frac{\delta}{a}$ as a function of redshift for different values of the BD parameter, $\omega$. }
	\label{fig:growth}
\end{figure}

The effective Newton's constant $G_{\rm eff}$ is close to $G$ at early times but it becomes larger at late times. This is because the scalar mode $\varphi$ gives an additional attractive force. We should note that this is opposite to the DGP self-accelerating solutions where the effective Newton constant is smaller due to the fact that the scalar mode is a ghost and it mediates a repulsive force. Solving the evolution equation numerically we obtained the growth factor $\delta/a$ as is shown in Fig.~\ref{fig:growth}. Due to the enhancement of Newton's constant, the growth rate is enhanced compared with the $\Lambda$CDM case.
\\

{\bf IV. Discussion}

By including a cubic order derivative interaction in the action for the Brans-Dicke theory, inspired by the DGP effective action and the Galileon model \cite{Nicolis:2008in}, we found a self-accelerating solution with no signs of ghost instabilities on small scales. This solution exists if the BD parameter is smaller than $-2$. Whenever the condition $H \gg \dot{\phi}/\phi > 0$ is satisfied initially, the resultant solutions are insensitive to the initial conditions for the scalar field. In order to get the observed acceleration, the parameter $M$ needs to be fine-tuned, $M \sim H_0$. A remarkable property of this cubic interaction is that it helps to recover GR at high energies. At high energies, this cubic interaction term ensures $\dot{\phi}/\phi \sim M \ll H$. This is a cosmological verision of the so-called Vainshtein mechanism \cite{Chow:2009fm}
\\
\indent The kinetic terms for small fluctuations on small scales, $d_t$ and $d_x$ are positive indicating that the model is free of ghost and tachyonic instabilities. The cubic interaction again plays a crucial role here. Without the cubic interaction, the fluctuations would be a ghost if $\omega < -4/3$, but the cubic interaction gives an additional contribution to the kinetic term for the fluctuations that changes the sign of the kinetic terms.
The sound speed can exceed one depending on model parameters and this super-luminality would impose constraints on the model parameters $\omega$ and $M$. We should note that there is the issue of super-luminal propagations in the DGP and Galileon model too, but this only occurs for large enough values of the parameter $|\omega|$ in our model. The full study of cosmological perturbations is necessary to ensure the stability of the model on horizon scales and this is an important open issue.
\\
\indent
At late times, gravity is strongly modified. This raises a question whether it is possible to evade the strong
constraints on the deviation from GR such as the solar system tests. Again the cubic interaction term comes to rescue. On small scales, including the non-linear term, the equation for $\varphi$ looks like
\be
 \frac{d_x}{a^2} \nabla^2\varphi + \frac{2}{a^4 M^2} \left[ (\nabla_i \nabla_j \varphi)^2 - (\nabla^2 \varphi)^2 \right]
 = -\frac{\rho}{\phi} \left(1 + \frac{P^2}{ M^2} \right) \delta, \nonumber
\ee
where $\varphi \ll 1$. This is the same equation as the one for the brane bending mode in the DGP model (see \cite{Koyama:2007ih}). For a static spherically symmetric source, it has been shown that GR is recovered on small scales $r<r_*$ where the Vainshtein radius $r_*$ is given by $r_*^3 \sim  r_g/M^2 d_x^2$, where $r_g$ is the Schwarzschild radius of the source. Since $M^{-1}$ is larger than the cosmological horizon scales, the Vainshtein radius is sufficiently large to recover GR and evade solar system constraints, but a small deviation from GR could be tested in future experiments.
\\
\indent
In summary the inclusion of the cubic interaction in the BD theory provides surprisingly rich phenomenology. It is then important to study the effects of the other higher order interactions discussed in \cite{Nicolis:2008in}. Moreover, since the Galilean symmetry is lost in the covariant version of the theory, there is no way to prevent higher order interactions from being generated and it is crucial to find a way to control these terms. This is related to the strong coupling problem, which constitutes yet another problem in the DGP model. This remains an open issue.

{\it Acknowledgments}:
FPS was supported by Funda\c{c}\~{a}o para a Ci\^{e}ncia  e a Tecnologia (Portugal)Ó, with the fellowshipÕs reference number: SFRH/BD/27249/2006. KK was supported by ERC, RCUK and STFC. KK would like to thank Justin Khoury and Wayne Hu for discussions.

\end{document}